\documentstyle[sprocl]{article}

\def\lb{\label}

\def\T{\mbox{\small \bf T}}

\def\be{\begin{equation}}
\def\ee{\end{equation}}
\def\bea{\begin{eqnarray}}
\def\eea{\end{eqnarray}}

\begin{document}

\title{PATH-INTEGRAL FOR QUADRATIC HAMILTONIAN SYSTEMS
AND BOUNDARY CONDITIONS}

\author{A. T. FILIPPOV , A. P. ISAEV}

\address{Bogoliubov Laboratory of Theoretical Physics, \\
JINR, Dubna, Moscow Reg., 141 980, Russia
\\E-mail: filippov@thsun1.jinr.ru, isaevap@thsun1.jinr.ru}


\maketitle\abstracts{
A path-integral representation for the kernel of the evolution
operator of general Hamiltonian systems is reviewed. We study
the models with bosonic and fermionic degrees of freedom.
A general scheme for introducing boundary conditions
in the path-integral is given. We calculate the path-integral
for the systems with
quadratic first class constraints
and present an explicit formula for the heat kernel (HK)
in this case. These
results may be applied to many quantum systems
which can be reduced to the Hamiltonian systems
with quadratic constraints (confined quarks, Calogero
type models, string and $p$-brain theories, etc.).
}

\section{Boundary conditions for path-integral.}

Let us consider a Hamiltonian system in the phase space ${\cal M}$ with
coordinates $Z_{A} = \{q_a, \, p^a, \, \psi_\alpha, \,
\overline{\psi}^{\alpha} \}$ where $q_a$ and $p^a$ are bosonic
coordinates and momenta $(a=1, \dots , N)$ while
$\psi_\alpha$ and $\overline{\psi}^{\alpha}$
$(\alpha = 1, \dots , K)$
are fermionic degrees of freedom. The action for this system can be
written in the form
\be
\label{1}
S = \int_{0}^{T} \, dt \, \left[ P^{A}(Z) \dot{Z}_A -
{\cal H}(Z) \right] \; ,
\ee
where $t \in [0,T]$ is an evolution parameter and ${\cal H}$
is the Hamiltonian. The variational principle,  $\delta S =0$,
gives the equations of motion,
$\dot{Z}_A = \Omega_{AB} \, \partial^B \, {\cal H}$.
$(\partial^{B} \equiv  \vec{\partial}^{B})$.
Here $\Omega_{AB} = -(-)^{AB} \, \Omega_{BA} =
\{ Z_A , \, Z_B \}$ is inverse to the matrix
$$
\Omega^{AB} =
(\partial^A \, P^B - (-)^{A + B + AB } \,
\partial^B \, P^A ) = -(-)^{A +B +AB} \, \Omega^{BA} \; .
$$
Note that the symplectic 2-form $\omega$
on the superspace ${\cal M}$ is defined by a slightly different supermatrix:
$\omega =
(\partial^A \, P^B - (-)^{AB } \, \partial^B \, P^A )
\, dZ_A \wedge dZ_B$.
Below we concentrate on the case of a constant supermatrix $\Omega^{AB}$.
Then we have
$P^{A} = \frac{1}{2} \, Z_B \, \Omega^{BA}$ and
the action (\ref{1})
can be represented in the form
\be
\lb{2}
S = \int_{0}^{T} \, dt \, \left[ \frac{1}{2} \,
Z_A \, \Omega^{AB} \, \dot{Z}_B -
{\cal H}(Z) \right] \; .
\ee

We will investigate the path-integral representation \cite{FaSl}
for the HK of the evolution operator $U$:
\be
\lb{3}
< Z^f | U(T,0) | Z^i > =
\int \, d \{ Z_A \} \, exp \{i (S + {\cal B}) \}
\ee
where $d \{ Z_A \}$ is a measure over the space of trajectories
in ${\cal M}$.
The boundary terms ${\cal B}$ depending on the initial and final points,
$Z^i$, $Z^f$, are determined by relevant boundary conditions.\\
{\bf Proposition 1.} {\it Boundary conditions can be
specified by the  boundary terms
in the path-integral (\ref{3}):
\be
\lb{4}
{\cal B} =  \frac{1}{2} \, \left[ Z^{i} \, \Omega \, P_i \, Z(0) -
Z^{f} \, \Omega \, P_f \, Z(T) \right] \; ,
\ee
where $Z^{f,i}_A$ are fixed supervectors in ${\cal M}$
and matrices $P_{f,i}$ are constant projectors ($P^2 =P$) in ${\cal M}$
of the rank $(N+K)$. These projectors should satisfy the conditions
$P^{\T} \, \Omega = \Omega \, (1-P)$, where $\T$ denotes a
super-transposition \cite{Ber,Wi}
$(P^{\T})_{AB} = (-)^{AB + A} \, P_{BA}$. Substituting the boundary
terms (\ref{4}) into (\ref{3}) is equivalent to
fixing the initial and final states of the system by the conditions:
$(1-P_i ) \, Z(0) = Z^i$,
$(1-P_f ) \, Z(T) = Z^f$.
}\\
{\bf Hint for proof.} Cosider
the equation $\delta (S +{\cal B}) = 0$ at the points
$t=0,T$. \\
{\bf Note.} Different choices of the projectors $P_{f,i}$
lead to different choices of the boundary states in (\ref{3}).
In this way one can write down the HK (\ref{3})
in coordinate, momentum or holomorphic representations
in unified form.

\section{Path-integral (\ref{3}) for
quadratic Hamiltonians.}

Now, consider systems with general quadratic
Hamiltonians
\be
\lb{5h}
{\cal H}(Z) = \frac{1}{2}
Z_A \, \left[ \Omega^{AB} \, {\cal A}^{D}_{B} \right] \, Z_D \; ,
\ee
where the supermatrix ${\cal A}$ depends on the evolution parameter $t$
and is independent of $Z_A$. The hermiticity of (5) requires that
${\cal A}^{\T} \Omega + \Omega {\cal A} =0$, i.e.
${\cal A} \in osp(2K|2N)$. With the Hamiltonian (\ref{5h})
the action (\ref{2})
acquires the form
\be
\lb{5}
S = \frac{1}{2} \,
\int_{0}^{T} \, dt \, Z \, \Omega \left[ \partial_t - A \right] \, Z \; .
\ee
\\
{\bf Proposition 2.} {\it
In the case of the Hamiltonian system with the
action (\ref{5})
and boundary terms (\ref{4}),
the path-integral (\ref{3}) gives the
following representation for the evolution operator $U$:
\be
\lb{6}
< Z^f | U(T,0) | Z^i > \simeq
Ber^{1/2} \left( \Omega \frac{1}{V^{+-}} \right) \,
\exp \left\{  - i \, S_{eff} (Z^f, \, Z^i) \right\} \; ,
\ee
$$
S_{eff} = \frac{1}{2} \, \left[ Z^i \Omega \frac{1}{V^{+-}} V^{++} Z^i
- 2 Z^i \Omega \frac{1}{V^{+-}} Z^f +
Z^f \Omega V^{--} \frac{1}{V^{+-}} Z^f \right]
$$
where $V=T \, exp \{ \int^{T}_{0} dt A(t) \}$,
$V^{\alpha \beta} = P^{\alpha}_f V P^{\beta}_i$ and
$P^{-}_{f,i} \equiv P_{f,i}$, $P^{+}_{f,i} \equiv 1-P_{f,i}$.
} \\
{\bf Proof.} Direct computation.

\section{Heat kernel for the
evolution operator of ``Discrete Strings''.}

Finally, consider the system with the action (\ref{5})
and interpret the Hamiltonian (\ref{5h}) as a linear combination
of the first class constraints. In this case, the elements of
the supermatrix $A$ play the role of Lagrange multipliers.
Note that the fermionic
string models can be described by the action
of the same type. For this reason we
call such systems as ``Discrete Strings'' (for details
see \cite{FI1,FI2,FI3,FI4}).
The Action with boundary terms, $S'=S+{\cal B}$
(where S and ${\cal B}$
are defined in (\ref{5}) and (\ref{4})),
is invariant under the gauge transformations
$$
Z \rightarrow F(t)Z \;\; , \;\;\;
A \equiv A_M e^M  \rightarrow F A F^{-1} + \dot{F}F^{-1} \Longleftrightarrow
V \rightarrow F(T) V F(0) \; ,
$$
where $F^{\T} \Omega F = \Omega$ (i.e., the gauge group
$G$ is a subgroup of $Osp(2K|2N)$). Matrices $e^M$ form a basis in
the algebra of $G$. The boundary
terms in $S'$ fix the gauge group parameters at the final and
initial points up to
residual gauge transformations:
$[ F(T,0), \, P_{f,i} ] = 0$.
Let us denote the corresponding groups by
$G_{f,i}$. Using the freedom of rotating the boundary states
$Z^{f,i} \rightarrow F(T,0)Z^{f,i}$, one can show that
the evolution operator (\ref{6}) depends only on the
coordinates of the moduli space
$\{ V \} / G_f \otimes G_i \equiv \overline{V}$,
which is an analogue of the Teichmueller space used in string
theories.

The standard procedure \cite{FaSl,FI3,FI5}
for quantizing constrained
Hamiltonian systems via path-integrals allows to obtain (after fixing
a gauge and representing Faddeev-Popov determinant
as an integral over ghosts $\{ B, \, C \}$) the expression
for the propagator of ``Discrete Strings'':
\be
\lb{7}
K_{f,i} =
\int \, d \{ \overline{V} \} \,
< Z^f | U(T,0) | Z^i > \cdot
< B^f, C^f | \widetilde{U}(T,0) | B^i , C^i> \; .
\ee
Here $d \{ \overline{V}\}$ is a measure over the moduli space
and the HK for the ghost evolution operator $\widetilde{U}$
is represented by the path-integral:
\be
\lb{8}
< B^f, C^f | \widetilde{U}(T,0) | B^i , C^i> =
\int \, d \{ B \} \, d \{ C \} \, exp \{i (S_{gh} + {\cal B}_{gh}) \}
\; .
\ee
The ghost action $S_{gh}$ and boundary terms ${\cal B}_{gh}$ have the form
$$
S_{gh} = \int^{T}_{0} dt \, B^N
(\delta^M_N \partial_t - \widetilde{A}^M_N )C_M \;\; , \;\;\;
{\cal B}_{gh} = B^i \widetilde{P}_i C(0) - B^f \widetilde{P}_f C(T) \; ,
$$
where ghosts $B^{N}$, $C_{M}$ realize the adjoint representations
of gauge group $G$
($\widetilde{A}$ is the adjoint analogue  of $A$)
and their parity is opposite to the parity of
the Lagrange multipliers $A_N$.
Constant projectors $\widetilde{P}_{f,i}$
are defined by the relations
$
(\widetilde{P}_{f,i})^{M}_{N} e^{N} =
P^{+}_{f,i} e^{M} P^{-}_{f,i} + P^{-}_{f,i} e^{M} P^{+}_{f,i}
$,
which impose some restrictions on the choice of the boundary
projectors $P_{f,i}$.
As in Proposition 1, the boundary terms ${\cal B}_{gh}$
fix the boundary states for ghosts:
$B(T,0)\widetilde{P}_{f,i} = B^{f,i}$,
$(1-\widetilde{P}_{f,i}) C(T,0) = C^{f,i}$.
\\
{\bf Proposition 3.} {\it The path-integral (\ref{8}) for the
ghost HK is equal to
\be
\lb{9}
< B^f, C^f | \widetilde{U}(T,0) | B^i , C^i> \simeq
\widetilde{Ber}(\widetilde{V}^{+-}) \, \exp \{ i \widetilde{S}_{eff} \} \; ,
\ee
$$
\widetilde{S}_{eff} =
\left[ (B^i - B^f \widetilde{V}^{--}) \frac{1}{\widetilde{V}^{+-}}
(C^f - \widetilde{V}^{++} C^i) - B^f \widetilde{V}^{-+} C^i \right] \; ,
$$
where $\widetilde{V} = T \exp \{ \int^{T}_{0} dt \, \widetilde{A}(t) \}$,
$\widetilde{V}^{\alpha \beta} =
\widetilde{P}^{\alpha}_f \widetilde{V} \widetilde{P}^{\beta}_i$
and
$\widetilde{P}^{-}_{f,i} \equiv \widetilde{P}_{f,i}$,
$\widetilde{P}^{+}_{f,i} \equiv 1-\widetilde{P}_{f,i}$.
} \\
{\bf Proof.} Direct computation.\\
{\bf Note.} We have to stress the following: if the operator
$\widetilde{V}^{+-}$ has zero eigenvalues, the right-hand side of
(\ref{9}) has to be modified. Namely, we have to remove the zero
eigenvalues from the superdeterminant
$\widetilde{Ber}$ and also multiply
the final expression  by the product of delta-functions of
corresponding zero-modes of the ghost variables $B,C$
(to remove infinities from $\widetilde{S}_{eff}$).

\vspace{0.1cm}

This work is supported in part
by the RFBR grant No. 97-01-01041 and by INTAS grant 93-127-ext.
We are grateful to CERN TH (A.P.I.) and INFN Torino (A.T.F.)
for kind hospitality and support.


\end{document}